# Discovery of potential collaboration networks from open knowledge sources

Nelson Piedra[1], Janneth Chicaiza[1], Jorge Lopez-Vargas[1] and Edmundo Tovar[2]

[1] Universidad Técnica Particular de Loja, Loja 1101608, Ecuador
[2] Universidad Politécnica de Madrid, Madrid 28660, Spain
`[nopiedra, jachicaiza, jalopez2]@utpl.edu.ec, etovar@fi.upm.es`

**Abstract.** Scientific publishing conveys the outputs of an academic or research activity, in this sense; it also reflects the efforts and issues in which people engage. To identify potential collaborative networks one of the simplest approaches is to leverage the co-authorship relations. In this approach, semantic and hierarchic relationships defined by a Knowledge Organization System are used in order to improve the system's ability to recommend potential networks beyond the lexical or syntactic analysis of the topics or concepts that are of interest to academics.

## 1 Introduction

Scientific publishing conveys the outputs of an academic or research activity, in this sense; it also reflects the efforts and issues in which people engage. The achievements made from academic and scientific work can help organizations to lead better resources management where they invest, and can help researchers to find peers or networks with whom to share development of a particular line of work. In this paper, the second topic is tackled.

To identify potential collaborative networks one of the simplest approaches is to leverage the co-authorship relations. In [1], methods of community detection were applied in order to build a network of co-authors. In [2, 3], areas of collaboration and common interests among researchers were identified through a social network approach.

Traditional methods applied to detect collaborative networks, as the abovementioned proposals don't use open sources of knowledge. Semantic and hierarchic relationships defined by a Knowledge Organization System (KOS) can help improve the system's ability to recommend potential networks beyond lexical or syntactic analysis of the topics or concepts that are of interest to academics. In this paper, this issue is undertaken through the semantic enrichment of the concepts associated with scholarly production. Open KOSs published under Linked Data Design Issues are used to find hidden or implicit relations between concepts that two researchers may be interested, so it is possible to identify potential collaboration communities.

## 2 Method

As a starting point, metadata of papers indexed in Scopus were collected from the dataset of linked data named "Open Data of Ecuador"[1] -one of the first to be published in Latin American about scientific production-, which provided the necessary scientific information.

Keywords of each paper were obtained querying the publications' dataset, from this data, the enrichment process was carried out. A popular dataset of linked data, named DBPedia, was used to add related concepts to each publications' subject.

Next, a query engine based-on Web Semantic technologies collected next data: i) papers related to concepts of a specific knowledge area, ii) authors and keywords of such publications, and iii) meta-concepts or top-concepts associated to keywords.

An inference mechanism based-on semantic relationship was enabled to discover communities that share an interest in a particular topic. The lexical and semantic analysis enabled to resolve next issues of grouping: i) similar terms as e-learning or elearning, ii) synonyms like "open educational resources" and" open learning materials ", and iii) acronyms as OCW which means "Open CourseWare".

## 3 Results

The proposal was validated with a subset of Scopus' papers with at least one author of any Ecuadorian affiliation. The analysis disclosed some features of the structure of the collaborative networks between authors, institutions and involved countries. Also, through the enrichment of concepts, main issues on which the researchers work were discovered.

---

[1] https://old.datahub.io/dataset/opendataec